 \documentclass[preprint,review,12pt]{elsarticle}

\usepackage{amsmath}
\usepackage{amsthm}
\usepackage{rotating}
\usepackage{multicol}
\usepackage{multirow}
\usepackage{textcomp}
\usepackage{lscape}
\usepackage{color}
\usepackage{bm}
\usepackage{graphicx}
\usepackage{epsfig}
\usepackage{adjustbox}
\usepackage{hyperref}
\usepackage[utf8]{inputenc}
\usepackage{url}
 \usepackage{lineno}

\newcommand{\figref}[1]{Fig.~\ref{#1}}

\newcommand{\tableref}[1]{Table~\ref{#1}}
\newcommand{\sectionref}[1]{\ref{#1}}

\graphicspath{ {./figures/} }

\biboptions{sort&compress}

\journal{Ultramicroscopy}

\begin{document}
\begin{frontmatter}



\title{Electron Backscattered Diffraction using a New Monolithic Direct Detector: High Resolution and Fast Acquisition}

\author[ucsb]{Fulin Wang}
\address[ucsb]{Materials Department, University of California Santa Barbara, Santa Barbara, CA 93117, USA.}

\author[ucsb]{McLean P. Echlin}

\author[ucsb]{Aidan A. Taylor}

\author[ucsb]{Jungho Shin}

\author[DE]{Benjamin Bammes}
\address[DE]{Direct Electron LP, San Diego, CA 92128 USA. }

\author[DE]{Barnaby D.A. Levin}

\author[cmu]{Marc De Graef}
\address[cmu]{Department of Materials Science and Engineering, Carnegie Mellon University, Pittsburgh, PA 15213, USA.}

\author[ucsb]{Tresa M. Pollock}

\author[ucsb]{Daniel S. Gianola\corref{cor1}}
\cortext[cor1]{Corresponding author. E-mail address: gianola@ucsb.edu}

\begin{abstract}

A monolithic active pixel sensor based direct detector that is optimized for the primary beam energies in scanning electron microscopes is implemented for electron back-scattered diffraction (EBSD) applications. The high detection efficiency of the detector and its large array of pixels allow sensitive and accurate detection of Kikuchi bands arising from primary electron beam excitation energies of 4 keV to 28 keV, with the optimal contrast occurring in the range of 8-16 keV. The diffraction pattern acquisition speed is substantially improved via a sparse sampling mode, resulting from the acquisition of a reduced number of pixels on the detector. Standard inpainting algorithms are implemented to effectively estimate the information in the skipped regions in the acquired diffraction pattern. For EBSD mapping, a speed as high as 5988 scan points per second is demonstrated, with a tolerable fraction of indexed points and accuracy. The collective capabilities spanning from high angular resolution EBSD pattern to high speed pattern acquisition are achieved on the same detector, facilitating simultaneous detection modalities that enable a multitude of advanced EBSD applications, including lattice strain mapping, structural refinement, low-dose characterization, 3D-EBSD and dynamic in situ EBSD.

\end{abstract}

\end{frontmatter}

\newpage

\section{Introduction}

Electron backscattered diffraction (EBSD) in the scanning electron microscope (SEM) has evolved to be a versatile and indispensable technique for phase identification, crystal orientation determination, and lattice strain measurement. Continued developments to advance the capabilities of EBSD generally follow one of several tacks.  On one front, acquiring EBSD patterns at high resolution is desirable for accurate measurements of lattice strain \cite{Troost1993_HREBSD,Vermeij2019_HREBSD,Wilkinson2006_HREBSD,Thomas2003_HREBSD,Britton2011_HREBSD}, structure refinement for low symmetry materials \cite{Prior2009,McCabe2015}, and cases such as transmission kikuchi diffraction (TKD) \cite{Keller2011,Trimby2012} where the diffracted intensities can be weak. Another direction focuses on advanced applications of EBSD, such as 3D serial sectioning \cite{Echlin2015,Echlin2020,Uchic2012,Burnett2016} and \textit{in situ} material testing and probing, which demand high speed pattern acquisition while preserving decent pattern quality for indexing sub-grain domains \cite{Singh2018_DeformedAl} and misorientation gradients \cite{Witzen2020,Polonsky2020}. In parallel, EBSD at low electron beam energies (accelerating voltages) is desirable for a number of reasons, including limiting charge in semi- or non-conducting samples \cite{Farrer2000}, reducing the electron-beam interaction volume for high spatial resolution mapping \cite{steinmetz2010,tripathi2019,Singh2018_DeformedAl}, and capturing enhanced band contrast, higher order bands or inelastic scattering effects \cite{Callahan2013,Winkelmann2008}. All these advancements would profit from high performance detectors that \textit{simultaneously} fulfill the requirements of resolution, speed and sensitivity.

The current state-of-the-art commercial EBSD systems use charge coupled device (CCD) or complementary metal–oxide–semiconductor (CMOS) sensor to capture the photons generated once diffracted electrons excite a phosphor screen. Usually, fiber optics or free space optics are used to project and focus an image of the backside of the phosphor onto a pixelated detector. Order of magnitude speed enhancements have been made over the past decade by upgrading the detectors, frequently with larger and larger so-called binning modes. Hardware binning of the detector works by grouping pixels into larger "effective" pixel sizes, or bins, and summing the aggregated signal. A major benefit of the binning approach is that shorter exposures can be used because the effective pixel size is larger. Although the angular resolution per binned pixel decreases with greater binning size, the binning approach has worked well due to parallel advancements in the sensitivity for band detection using Hough-based indexing algorithms (used in most commercial EBSD software packages) \cite{lassen1992_Hough,Zambaldi2009}. Furthermore, dynamical EBSD pattern simulation \cite{EMsoft5} and the resulting EBSD dictionary indexing methods have proven to be exceptionally sensitive for accurate band detection on detectors with a small number of pixels and with low signal-to-noise ratio (SNR) \cite{Ram2017,Chen2015,Jackson2019,Singh2016}. Speed improvements and memory requirements for on-the-fly indexing, particularly for indexing high symmetry materials, have been realized with spherical EBSD indexing \cite{Lenthe2019}. 

Further improvement in the performance of the phosphor-based EBSD systems is limited owing to their detection approach, specifically in terms of pattern quality (resolution and contrast) and acquisition speed. This form of indirect detection technique is fundamentally inefficient as signal degradation and information-sharing between pixels can occur at several points along the system. Achieving even higher frame rates is challenging because of the inadequate SNR in each high-speed frame. These challenges can be overcome by direct detection of electrons, where electrons are directly converted to electrical signals on a pixelated detector. There are primarily two types of direct detectors. The first type is the hybrid pixel array detector (PAD), where a thick sensing layer (Si or CdTe) is bump-bonded to the separate readout electronics. The pixel size on the thick sensor is relatively large, e.g. 55 $\mu$m on the Medipix series \cite{medipix} or 150 $\mu$m on EMPAD \cite{tate2016,Nguyen2018}. The large sensing volume of each pixel allows high energy electrons to be fully sensed, and yields appealing features including high SNR for single electron hits and high dynamic range \cite{tate2016}. The second category of detectors are known as monolithic active pixel sensors (MAPS), where the sensor and readout electronics are integrated in a single semiconductor wafer. The sensing layer is very thin, for example 8 $\mu$m \cite{MILAZZO2005152,MCMULLAN2014156}. This limits the lateral spread of electrons in the sensor, and hence allows a large number of pixels to be incorporated in the sensor with small pixel size. Compared to PAD, MAPS provides higher resolution based on the reduced pixel size but lacks the wide dynamic range. 

The development of modern direct detectors has led to breakthroughs in transmission electron microscopy (TEM) for techniques such as single particle cryo-electron microscopy (using MAPS) \cite{bammes2012,li2013electron} and 4D-STEM (using both MAPS and PAD) \cite{muller2014,ozdol2015,krajnak2016,yang2017,jiang2018,ophus2019,zhang2020,zhou2020}. Direct detectors have also been employed for EBSD applications and their utility have been demonstrated in terms of acquiring high quality diffraction patterns, as well as the application of energy filtering to EBSD \cite{Wilkinson2013,vespucci2015}. The first implementation of direct detectors for EBSD by Wilkinson et al. \cite{Wilkinson2013} used a CMOS sensor that is backthinned (i.e. a MAPS) to improve the detection of low-energy electrons in SEM, and several subsequent studies used the Medipix2 detector (i.e. a PAD) \cite{vespucci2015,mingard2018}. For practical EBSD mapping, the acquisition rate of EBSD patterns on the current direct detection cameras is relatively slow (\tableref{table_detectors}), which was ascribed to limitations in the read-out system, not the fundamental sensor performance \cite{mingard2018}. Taken as a whole, this suggests that fully profiting from the strengths of direct detection for EBSD applications hinges on continued improvements in direct detection systems.


Here, we report the implementation of a MAPS direct detector that is optimized for the SEM environment. The high sensitivity and large array of pixels (2k $\times$ 2k) of the detector are highlighted, both of which are essential for detecting diffraction features at high angular resolution and especially at low electron beam energy (4 keV to 28 keV). We demonstrate substantially improved scan speed for EBSD mapping via a sparse sampling mode, achieved by reading and saving a reduced number of pixels on the detector. Using standard algorithms to in-paint the skipped regions in the acquire diffraction patterns, the data in the fast-scan mode can be indexed by spherical indexing method at a tolerable indexing fraction and accuracy up to $\sim$ 6000 scan points per second. 

\bgroup
\def\arraystretch{1}
\begin{table*}[htb]
\caption{Specifications of representative EBSD detectors in SEM. The EDAX Velocity\textsuperscript{TM} is an indirect detection camera, using a CMOS sensor coupled to a scintillator, targeting high-speed EBSD mapping. The rest are direction electron detection cameras. The frame rates (fps: frame per second) are with respect to full resolution. The EBSD scan speeds (pps: points per second) are the highest speed reported with binning modes or sparse sampling. Binning modes retain the same active sensor area, but with large effective pixels. Sparse sampling modes reduce the total number of active pixels, and therefore active sensing area. Thus binning and sparse sampling modes are not directly comparable.}
\centering
\begin{adjustbox}{width=\textwidth}
\begin{tabular}{c|c|c|c|c|c}\hline
Detection & Detector & Pixels & Pixel Size ($\mu$m)  & Frame Rate & EBSD speed \\ \hline
Indirect & EDAX Velocity\textsuperscript{TM} & 640$\times$480 & 60-70  & - & 3000, 4500\\ \hline
Direct, PAD & EDAX Clarity\textsuperscript{TM} & 512$\times$512 & 60-70  & - & 85 \\
Direct, PAD & Medipix 2 & 256$\times$256 & 55 & 1400 \cite{mexipix} & 19 \cite{mingard2018} \\
Direct, MAPS & Modified CMOS \cite{Wilkinson2013} & 1024$\times$1024 & 20 & 28 & - \\ \hline
Direct, MAPS & DE-SEMCam & 2048$\times$2048 & 13 & 281 & $\sim6000$ \\\hline
\end{tabular}\end{adjustbox}\label{table_detectors}
\end{table*}\egroup

\section{Methods}

    \subsection{Integration of the Direct Detector in SEM}
    
\begin{figure*}[htb]
\centering
\includegraphics[width=0.8\textwidth]{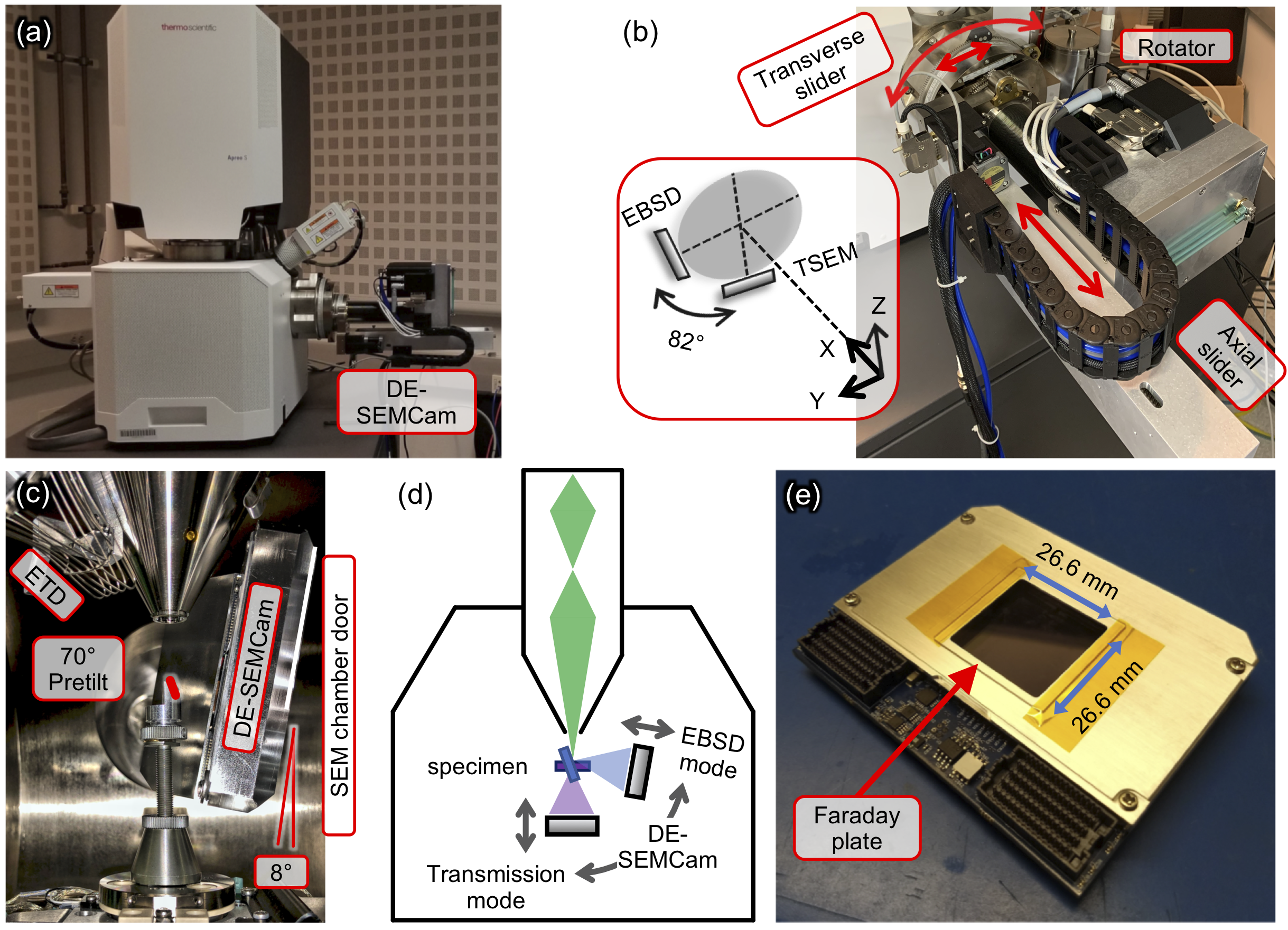}
\caption{(a) DE-SEMCam installed on the Thermo Scientific\textsuperscript{TM} Apreo-S SEM. (b) View of the DE-SEMCam from the right side of the SEM. The schematic shows the rotation and translation axes of the positioning system. (c) The chamber view from the left side of the SEM. (d) Schematic of the two operation modes, TSEM and EBSD, of the DE-SEMCam in the SEM chamber. (e) A view of the front face of the direct detector.}
\label{sem_view}
\end{figure*}

The camera we evaluated here, DE-SEMCam, was manufactured by Direct Electron LP (San Diego, CA USA) using a custom MAPS (called a Direct Detection Device (DDD\textsuperscript{\textregistered})), which is designed for detection of low-energy electrons, based on an earlier-prototype low-energy-optimized direct detector designed for low energy electron microscope (LEEM) or photo electron emission microscope (PEEM) \cite{tromp2015}. It has a much smaller pixel size of 13 $\mu$m and features a large total number of pixels in comparison to a hybrid PAD (\tableref{table_detectors}). The camera is installed on the Thermo Scientific\textsuperscript{TM} Apreo-S SEM. A mechanical positioning system was designed for the DE-SEMCam that provides both translational and rotational degrees of freedom of the detector within the SEM chamber. In the coordinate system shown in \figref{sem_view} (b), the X-axis is the axial direction pointing into the SEM chamber from the right flange, the Y-axis is parallel to the detector plane and the Z-axis is normal to the detector plane. The camera is inserted and retracted along the X-axis. Rotation around the X-axis transitions the camera between the EBSD mode (Z-sxis 8$^{\circ}$ above the horizontal plane) and the transmission in SEM (TSEM) mode (Z-axis pointing parallel to the electron column path). In the TSEM mode, on-axis diffraction \cite{FUNDENBERGER2016,BRODU2017} and diffraction contrast imaging \cite{CALLAHAN2018,STINVILLE2019} are available, as well as TEM-type diffraction mapping modes such as 4D-STEM. For both the EBSD and TSEM modes, the sample-to-detector distance (camera length) can be adjusted by translation along the Z-axis.

In the present study, we focus on the application of the DE-SEMCam for EBSD. The geometry of the camera in EBSD mode is similar to conventional EBSD systems. The inclination angle between the detector and the horizontal plane is 8$^{\circ}$, as shown in \figref{sem_view} (c). The geometric parameters of the camera are determined using the Efit program in EMsoft software package \cite{EMsoft5} by matching an experimental EBSD pattern with the theoretically computed pattern. Using this method, the sample-to-detector distance is resolved to be 22.402 mm and the pattern center coordinates are $(x^{*}, y^{*}, z^{*}) = (0.4950, 0.7005, 0.8414)$ using the EDAX/TSL convention. The DE-SEMCam has a sensor size of 26.6 mm, which covers a solid angle of $\sim61^\circ$ when using the parameters described here. Different geometric parameters can be realized by adjusting the inclination angle of the camera and the sample-to-detector distance using the mechanical positioning system. The clearance envelope is defined by the pole-piece of the SEM and the specimen stage, which is especially relevant when small sample-to-detector distance is needed to cover a large solid angle.

    \subsection{Acquisition modes and data format} 

The full-frame resolution of the detector is 2048$\times$2048 with 13 $\mu$m pixel size and a maximum readout speed of 281 fps. The imaging resolution is significantly higher than those on conventional EBSD detectors (\tableref{table_detectors}). The frame rates can be further increased by reading from a reduced number of rows, because the readout time of one frame depends on the number of pixel rows being addressed on the detector. We employ a sparse sampling and compressive readout mode on the DE-SEMCam, termed Arbitrary Kernel Row Addressing (AKRA) \cite{bammes2019}, where the user specifies an arbitrary number and position of the rows to be addressed on the detector. This is based on the unique architecture of the DE-SEMCam sensor and is not feasible on many other CMOS-based detectors. The pixel rows on DE-SEMCam are grouped four at a time into kernel rows, each of which has a unique address. The top kernel row (group of four pixel rows) is denoted with address 0, the next kernel row is address 1, and so on until the last kernel row which is address 511. During readout of each frame from a CMOS-based detector, kernel rows are read out sequentially. Sub-array readout sets the range of kernel rows to read out, but it is only possible to read out a contiguous region from the sensor. Therefore, sub-array readout on conventional CMOS-based detectors results in a smaller solid angle being captured. In contrast, DE-SEMCam's AKRA-mode enables users to read fewer kernel rows (resulting in an increase in frame rate), while spreading the kernel rows across the entire sensor area so that the solid angle captured by the detector remains constant. When DE-SEMCam's AKRA-mode is enabled, a table of kernel row addresses is loaded in the camera's internal memory. During readout of each frame in AKRA-mode, kernel row addresses in the AKRA table are sequentially fetched and used to read out only the specific kernel rows specified in the AKRA table. This processing is accomplished internally in the camera's firmware and does not cause any reduction in speed.

In this study, we distribute the actively read-out kernel rows sparsely over the whole detector with equal distance, as shown in \figref{akra_schematic} (a). With increasing distance between the addressed rows, the frame rate of the detector increases as shown in \tableref{t_akra}, yet the solid angle range of detection is not compromised. For instance, if we use a sampling step of 8, then we would load the following addresses into the AKRA table: 0, 8, 16, 24, ..., 496, 504. In this case, for every kernel row (4 pixel rows) that is read out, 7 kernel rows (28 pixel rows) would be skipped (not read). The detecting area of the sensor therefore decreases to 12.5\% and only 256 pixel rows (instead of 2048 pixel rows) are read out. This results in a significant increase in frame rate, in this case increased to 2200 fps.

When the readout of the detector is synchronized to the electron beam scanning, the frame rate of the DE-SEMCam in AKRA mode directly equals the EBSD scan speed (points per second). The reduced number of active rows using AKRA has two major consequences: (1) the fast readout rate reduces the number of electrons collected on the detector due to the short exposure time and (2) a sparsely sampled diffraction pattern on the detector leads to a physically reduced detecting area. Thus, for a fixed current density of electrons exiting the sample, the detected electron dose is proportional to the product of the active detecting area and the exposure time. A normalized dose for each of the AKRA modes is computed with reference to the electron dose that exists during full-frame detection and is listed in \tableref{t_akra}. For instance, a diffraction pattern obtained using the AKRA mode at 5988 fps leads to a detector dose of only 0.2\% of that of a full-framed pattern, emphasizing the need for high detection efficiency. 

Since the geometry of the DE-SEMCam in EBSD mode is approximately the same as the conventional system (solid angle detection areas were designed to be equal), we can also normalize the electron dose on the conventional camera at different binning sizes by that of the DE-SEMCam at full-frame. In the case of the conventional EBSD detector, the detecting area remains as the full area of the detector at larger binning modes; however, the exposure times decrease. Thus, the high frame rate of the DE-SEMCam operating in the AKRA mode is achieved at the expense of the detected electron dose when compared to full-frame detection. As will be demonstrated in the present work, however, an extremely high electron detection sensitivity exists for these limited number of active pixels, allowing for the fast collection of indexable EBSD patterns.  Moreover, these features provide flexibility to the user to either selectively address pixels on the detector or acquire the full pixel array, for high-speed or high-resolution EBSD applications, respectively.


\bgroup
\def\arraystretch{1}
\begin{table*}[htb]
\caption{Detection parameters of the DE-SEMCam for EBSD scans with different AKRA modes. The parameters for EBSD scans using a conventional (Con.) EDAX Hikari Plus detector at different binning modes are also shown. The highest possible frame rates are shown for each mode. The sampling step is the number of kernel rows from the start of the previous addressed kernel rows to the start of the next. The speedup factor is with respect to the full-frame frame rate of 281 fps. The normalized electron dose is calculated as the product of the active pixel detecting area and dwell time (1/frame rate), and normalized to the full-frame 2048$\times$2048. }
\centering
\begin{adjustbox}{width=\textwidth}
\begin{tabular}{c|l|c|c|c|c|c}\hline
 & Resolution & Sampling Step & Detecting Area (\%)  & Frame Rate (fps) & Speedup & Normalized Dose \\ \hline
\multirow{5}{12pt}{\rotatebox[origin=c]{90}{DE-SEMCam}} & 2048$\times$2048  &  1  &  100 &  281 & - & 100\\ 
                                            & 2048$\times$256 & 8  &  12.5 &  2200 & 7.83 & 1.60\\ 
                                            & 2048$\times$128 & 16  &  6.25 &  4237 & 15.1 & 0.415\\ 
                                             & 2048$\times$104 & 20  &  5.08 &  5154 & 18.3 & 0.277\\ 
                                             & 2048$\times$88 & 24  &  4.30 &  5988 & 21.3 & 0.202\\ \hline
\multirow{3}{12pt}{\rotatebox[origin=c]{90}{Con.}} & 480$\times$480 & 1$\times$1 binning   &  78.5 &  118 & -  & 187\\ 
                                             & 120$\times$120 & 4$\times$4 binning   &  78.5 &  661 & 2.35  & 33.4\\ 
                                             & 30$\times$30 & 16$\times$16 binning  &  78.5 &  1550 & 5.52  & 14.2\\\hline
\end{tabular}\end{adjustbox}\label{t_akra}
\end{table*}\egroup

    \subsection{Post-collection processing of the sparsely sampled patterns}

The EBSD patterns captured in the AKRA mode are a sparse sampling of the full pattern on the detector with skipped regions, resulting in a raw image that is compressed in one dimension. In order to index the sparsely sampled diffraction patterns, we used inpainting to restore the skipped regions based on the information collected by the active pixels.

\begin{figure*}[htb]
\centering
\includegraphics[width=0.8\textwidth]{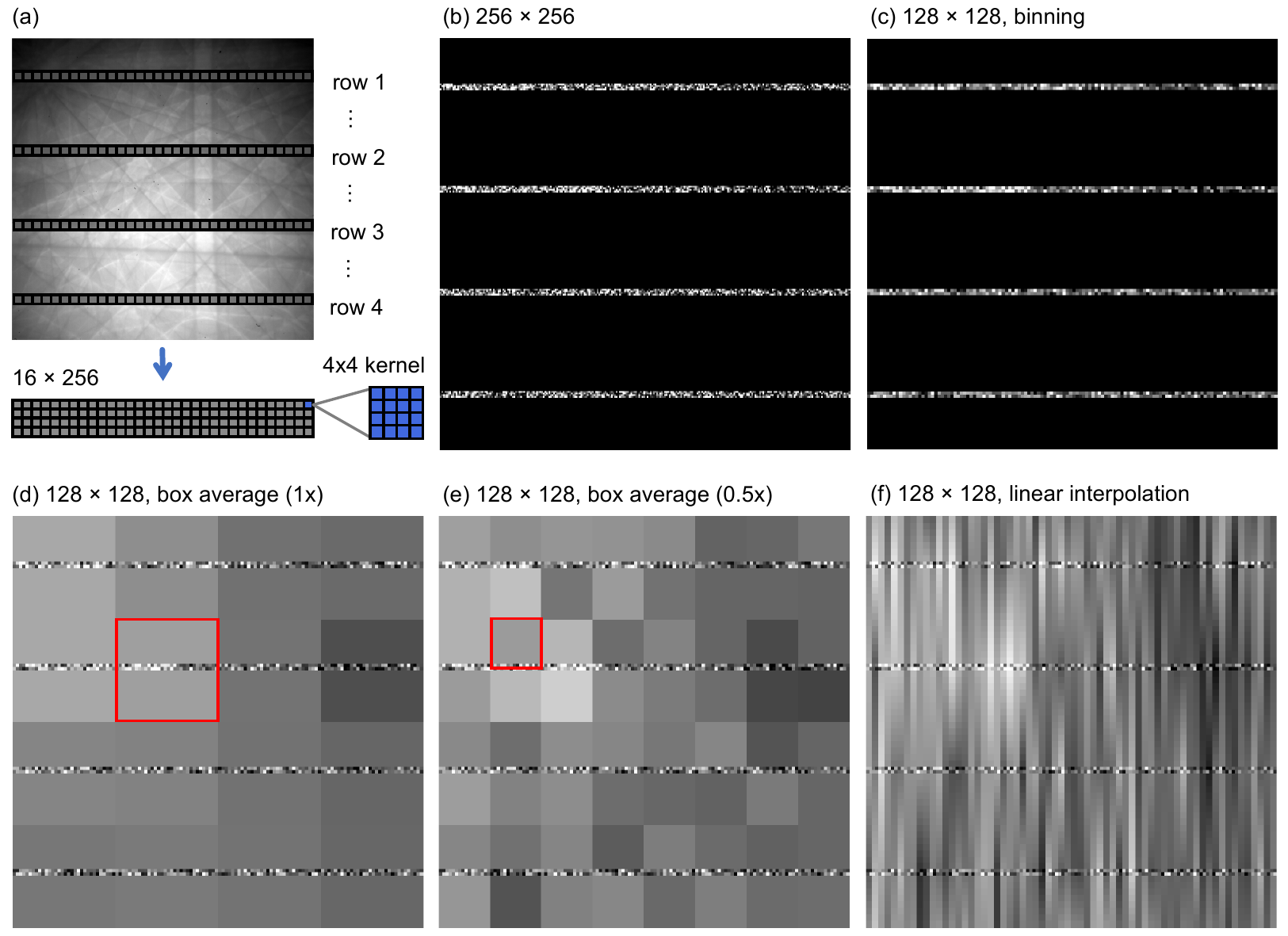}
\caption{Sparse sampling of EBSD patterns and post-collection processing methods. The schematic in (a) shows the sparse sampling of a EBSD pattern by addressing 4 equally spaced kernel rows. In this example a full sensor resolution of 256$\times$256 is used for demonstration and ease of visualization. A sampling step of 16 kernel rows (64 pixels) is applied, resulting in a compressed AKRA image with pixel dimension of 16$\times$256, with 4$\times$4 pixels in a kernel. (b) The sparsely sampled pattern, as produced by restoring the spatial positions of the kernel rows and filling the inactive rows using global inpainting of zero intensity (black pixels). (c) is produced in the same way as (b), but with a software binning of 2$\times$2. (d) to (f) are produced using location sensitive inpainting methods including box averaging (box size is 1$\times$ and 0.5$\times$ AKRA sampling step size) and linear interpolation. The red boxes in (d) and (e) highlight the size of the box relative to the sampling step or inactive row spacing.}
\label{akra_schematic}
\end{figure*}

In this study, we adopted image processing and inpainting routines that reflect either simple computations or commonly-adopted processes to highlight the intrinsic nature of the detected diffraction patterns, thereby allowing for a more apt comparison of detection schemes. More advanced methods will likely improve the quality of EBSD mapping results, but we do not include these here. Specifically, each AKRA pattern was initially processed with a background subtraction, which is computed by averaging a large number of patterns in the complete data set. Background subtractions are almost universally applied to conventionally EBSD data, except for unique cases such as single crystalline materials. Each pattern was then software binned by a factor of 2 to improve the SNR. A simple inpainting method is blanket replacement of inactive pixels with a fixed value, such as zero intensity shown as black in \figref{akra_schematic} (b), or gray pixels whose intensity is the average of the values of all the active AKRA pixels. These fixed value inpainting methods are not easily indexed using the method used here. In order to better describe the local diffraction features, inpainting is performed based on the pixels in the neighboring AKRA rows. 

One method to accomplish this is a box averaging approach where the box edge size is proportional to the spacing between active AKRA rows. The box averaging approach vertically expands the measurements at the active pixel regions, compensating for the anisotropic sensing (compression along the vertical detector axis) that the AKRA mode imprints on the EBSD patterns. \figref{akra_schematic} (d) and (e) show examples of box averaging with box sizes = 1$\times$ and 0.5$\times$Sampling Step, respectively. Another method that was employed is linear interpolation between the active AKRA rows, where the values of the inactive rows are interpolated from the previous AKRA row to the next, as shown in \figref{akra_schematic} (f). All of the inpainting methods presented are intentionally straightforward, however more sophisticated methods could be applied, such as the Telea algorithm \cite{telea2004} that is included in the OpenCV library \cite{opencv}. 

    \subsection{EBSD and Indexing} \label{sec:Methods:XC}

We first used single crystal Si to examine the EBSD pattern quality as a function of the accelerating voltage of SEM, and then used polycrystalline Ni specimen to evaluate the efficacy of fast scan in AKRA mode for EBSD mapping. The single crystal Si EBSD patterns were acquired with the microscope operating in spot mode, and on the DE-SEMCam with a full-frame resolution of 2048$\times$2048 and a 13 $\mu$m pixel size, as well as on a TSL/EDAX EBSD Hikari Plus detector on Thermo Scientific\textsuperscript{TM} Versa 3D SEM at a full resolution of 480$\times$480 and a 67.5 $\mu$m pixel size. The acquired patterns were only adjusted by histogram equalization when compared to each other.

EBSD mappings were performed on a polished polycrystalline Ni specimen using the DE-SEMCam and the TSL/EDAX EBSD Hikari Plus detector. The respective microscopes were operated at 10 keV and 20 keV accelerating voltages with an electron beam current of 32 nA. The scan area is 200$\times$200 $\mu\text{m}^2$ with a 1 $\mu$m step size, consisting of 40,000 scan points. EBSD patterns were recorded with the DE-SEMCam at full-frame and the different AKRA modes shown in \tableref{t_akra}. The detector was synchronized to the electron beam using the DE-FreeScan scan controller (Direct Electron LP, San Diego, CA USA), which receives a TTL trigger signal from the DE-SEMCam at the end of each frame acquisition. With the conventional EDAX EBSD detector, three binning modes were used, and the camera gain and exposure were adjusted to achieve the highest possible frame rates at each binning mode. The exposure times for the EDAX EBSD detector at 10 keV are 8.5 ms (1$\times$1 binning), 1.45 ms (4$\times$4 binning), and 0.6 ms (16$\times$16 binning). At 20 keV the exposure times are 4.95 ms (1$\times$1 binning), 1.45 ms (4$\times$4 binning), and 0.6 ms (16$\times$16 binning). All the acquired EBSD patterns were enhanced with an standard background subtraction (background generated from the average of a large number of EBSD patterns). The AKRA images on the DE-SEMCam were additionally processed by inpainting.

The processed EBSD patterns were all indexed using EMSphInx \cite{Lenthe2019} with a bandwidth of 128. The spherical indexing method in EMSphInx uses the detector geometrical parameters to back-project the experimental pattern onto the Kikuchi sphere. The spherical harmonic transform is used to evaluate the cross-correlation between a back-projected experimental pattern and the spherical master pattern. A similarity metric is calculated in Euler space. The maximum cross-correlation (XC) value represents the optimal similarity and hence determines the most likely crystallographic orientation of the experimental pattern. Since the cross-correlation is performed on the entire pattern, the XC values describe the certainty of the solution within all of orientation space. Similar to the dot-product in the dictionary-based EBSD indexing method \cite{Chen2015,Jackson2019}, the XC values are in the range of 0 to 1.


\section{Results}

    \subsection{High resolution EBSD patterns over a wide range of accelerating voltages} \label{wide_keV_range_Si}

\begin{figure*}[htb!]
\centering
\includegraphics[width=1\textwidth]{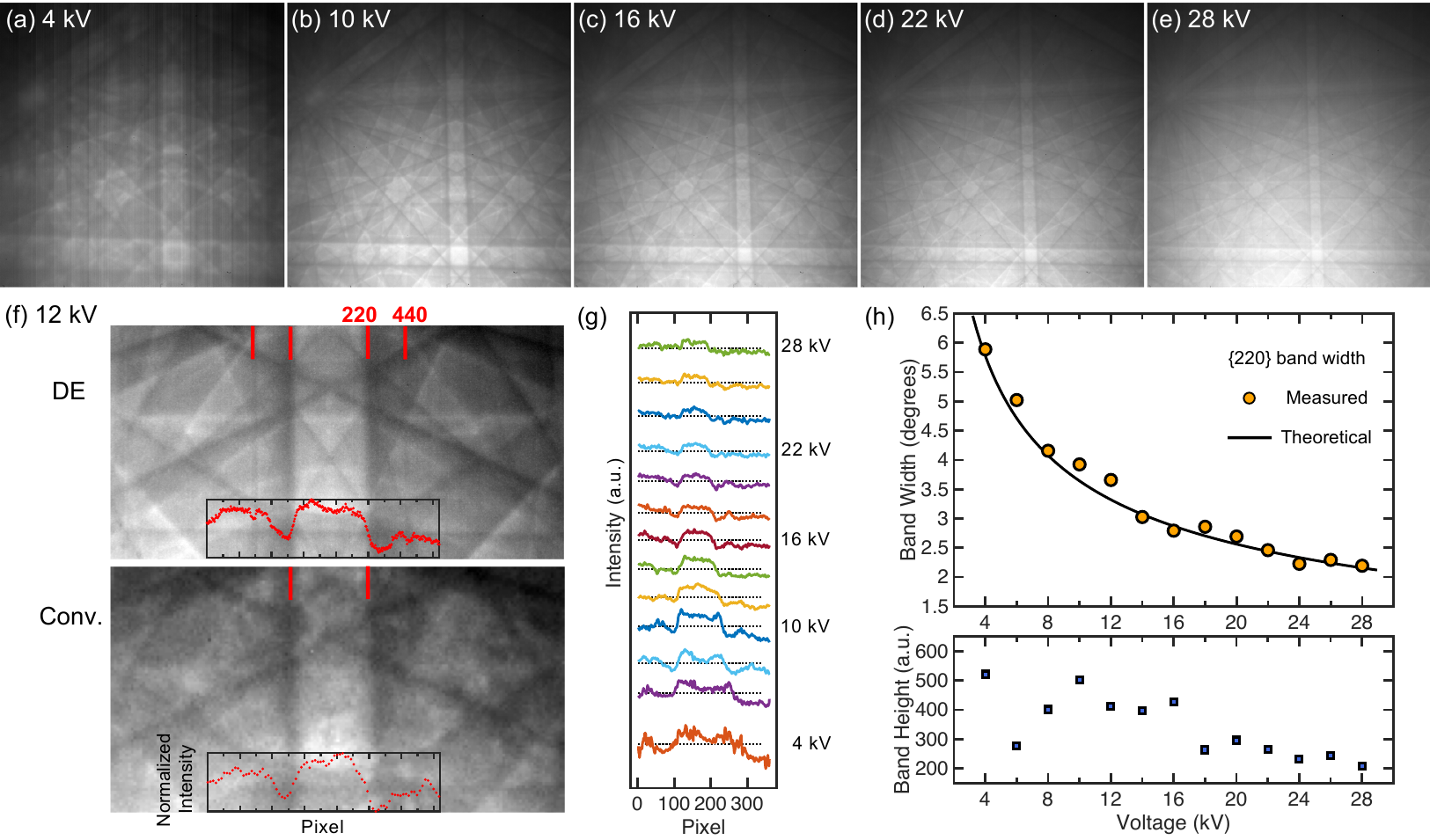}
\caption{(a-e) EBSD patterns from single crystal Si were acquired using the DE-SEMCam (full-frame 2048$\times$2048 pixels with pixel size 13 $\mu$m) at a range of electron beam accelerating voltages (keV), a 6.4 nA beam current and 1 s exposure. (f) Magnified regions of the EBSD patterns acquired using the DE-SEMCam and conventional (Conv.) EDAX EBSD detector (operated at full-frame 480$\times$480 pixels with pixel size 67.5 $\mu$m), at 12 keV, 13 nA beam current and 1 s exposure. The insets in (f) are line profiles extracted across the \{220\} Kikuchi band. The intensities on the two detectors are normalized to the respective minimum and maximum values. The line profiles across the \{220\} band are plotted from 4 keV to 28 keV in (g). The dotted lines indicate the average level for each profile. The band widths and band heights are measured from (g) and plotted in (h) as markers. The solid line is the theoretical relationship between the Si \{220\} band width and voltage.}
\label{highres_pat}
\end{figure*}

The high resolution and sensitivity of the DE-SEMCam is demonstrated using EBSD patterns obtained from primary beam energies ranging from 4 keV to 28 keV. The DE-SEMCam consistently produces Kikuchi diffraction features at all the tested voltages as shown in \figref{highres_pat} (a)-(e), readily capturing diffraction details such as band broadening and high-order reflections. A magnified view around the \{220\} band at 12 keV is shown in \figref{highres_pat} (f) and is compared with that acquired on a conventional indirect EBSD detector. Rich and well-defined diffraction features are captured by the DE-SEMCam. A line profile perpendicular to the \{220\} band shows the \{220\} band edges as well as the high-order \{440\} reflections that are clearly resolved as a sharp drop in the intensity profile. In comparison, the same pattern from the conventional detector reveals only the \{220\} band edges but not the high-order reflections. Note that the intensity drop across the \{440\} edge spans 17 pixels in the DE-SEMCam image, and only 3 to 4 pixels in the conventional detector image. Both the high detection efficiency and the small pixel size of the DE-SEMCam are essential for accurately detecting the Kikuchi band edges.

From the EBSD patterns at varying voltages, the intensity profiles across the \{220\} band were measured (\figref{highres_pat} (g)). The Kikuchi band manifests as a plateau of higher intensity in the profile, with the left edge at around the 100th pixel and the right edge in the range between the 200th and the 300th pixel. A decrease in band width with increasing voltage is clearly revealed. The bandwidths at varying voltages were further measured from the profiles as the full width at the half maximum of the plateau. Using the pixel size of 13 $\mu$m and sample-to-detector distance of 22.402 mm, the bandwidths are converted to angular distances and are plotted against the respective voltages in \figref{highres_pat} (h). The theoretical Bragg angle of the \{220\} band in Si as a function of accelerating voltage is calculated from $2d\sin{\theta_{B}}=\lambda$, where $d$ is the interplanar spacing of the \{220\} planes, and $\lambda$ is the relativistic electron wavelength depending on electron energy (keV). All the measured \{220\} band widths are consistent with the theoretical values across the voltage range of 4 keV to 28 keV. The band heights as a function of voltage were also quantified as the difference in intensity between the band edge (the trough) and the band plateau in \figref{highres_pat} (h). A larger height indicates higher contrast at the band edge. While the apparent large band height at 4 keV is a result of noise and therefore does not reflect good contrast, the optimal contrast is observed between 8 keV and 16 keV accelerating voltage. 

The accurate detection of the Kikuchi band width and the ability to resolve high-order diffraction features suggest a promising avenue for applications such as lattice strain measurement using high resolution EBSD (HR-EBSD) \cite{Troost1993_HREBSD,Vermeij2019_HREBSD,Wilkinson2006_HREBSD,Thomas2003_HREBSD,Britton2011_HREBSD} or phase refinement, e.g. differentiating pseudo symmetric phase variants \cite{brewer2010,lenthe2019ps}. Not only does the DE-SEMCam have high electron detection sensitivity and small pixel sizes, but it also has enhanced contrast at 8-16 keV where Kikuchi bands are wider than at higher voltages, leading to improved accuracy of lattice strain measurements. Furthermore, mapping of deformed microstructure with large orientation gradients and small grain sizes will benefit from lower keV EBSD measurements that reduce the interaction volume and hence increase the EBSD spatial resolutions \cite{Singh2018_DeformedAl}.

    \subsection{High Speed EBSD Mapping using the DE-SEMCam}

The full-frame images on the DE-SEMCam, highlighting the advantages of accurate and high resolution EBSD, can be acquired at a maximum frame rate of 281 fps for 2048$\times$2048 pixels. Much higher speeds can be achieved when the detector operates in the AKRA mode where a fraction of the detector pixel rows, rather than the full frame, are addressed. In the following, EBSD patterns and the indexing results from fast-scan experiments are presented. All results are presented in comparison to the full-frame DE-SEMCam scans and the scans with a conventional EDAX EBSD detector operating at different binning modes.
   
        \subsubsection{Patterns in the fast-scan mode}

\begin{figure*}[htb!]
\centering
\includegraphics[width=\textwidth]{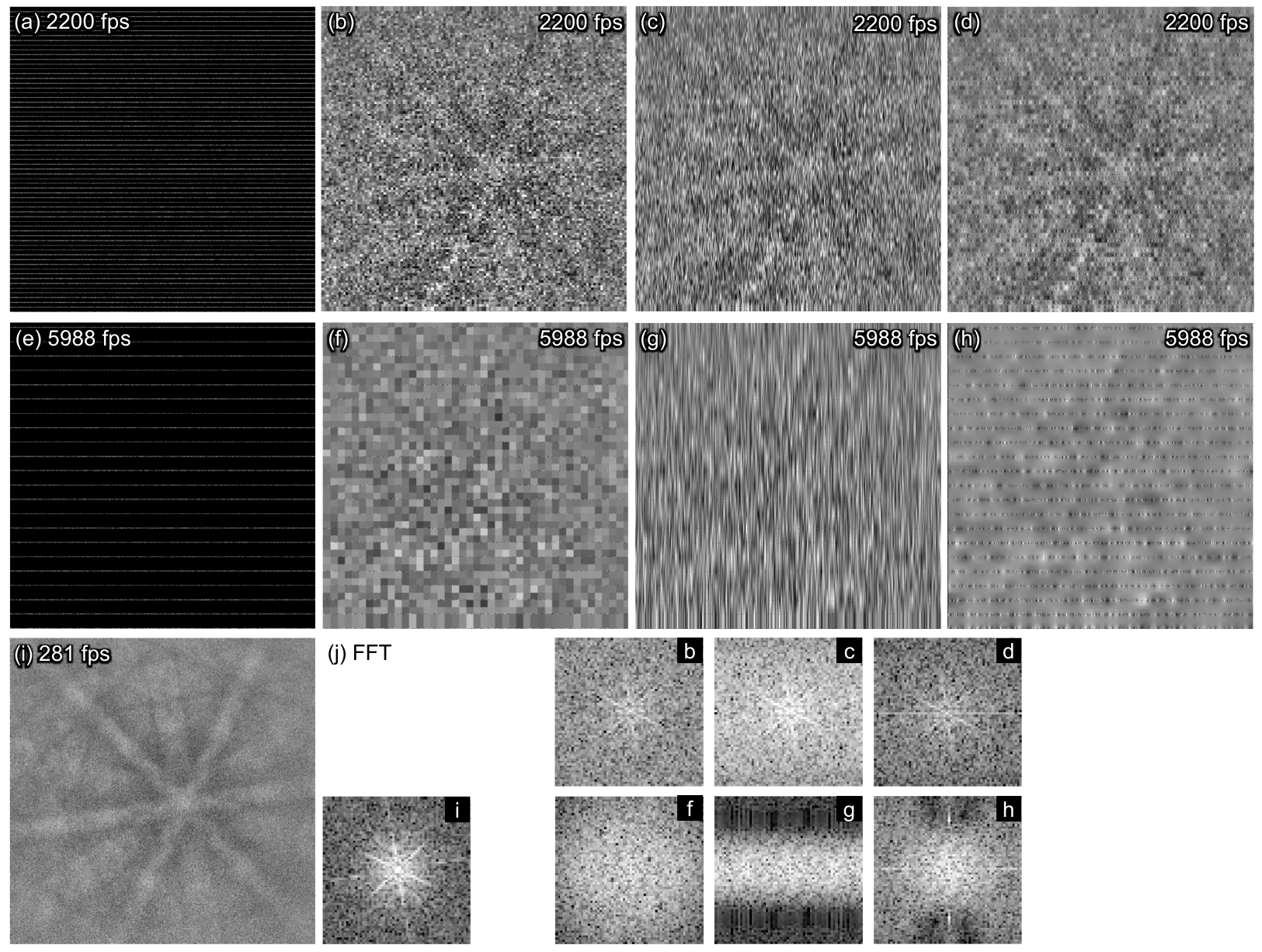}
\caption{AKRA patterns from DE-SEMCam processed using different inpainting methods are shown: (a \& e) inpainting with zero intensity, (b \& f) box averaging with a size of 0.5$\times$sampling step, (c \& g) linear interpolation, and (d \& h) the algorithm of Telea \cite{telea2004}. A full-frame pattern captured at 2048$\times$2048 is presented in (i). All the patterns are single frame images acquired at the indicated frame rate and from the same spot on the Ni specimen. The electron beam voltage was 10 keV and the beam current is 32 nA. A background image was subtracted from all the patterns. The FFT of the selected patterns are shown in the panel in (j). Each set of patterns: (a) to (d), (e) to (h), and the set of FFT images were adjusted to be within the same contrast range.}
\label{akra_patns}
\end{figure*}        

The raw patterns acquired in the AKRA modes have a reduced size in one dimension, and more importantly, reduced information. The sparsity of the detecting area can be readily observed in the images in \figref{akra_patns} (a) and (e) that are in-painted with zero intensity values. The resulting patterns appear as undulating lines of signal on a constant background. The Kikuchi bands are vaguely discernible by eye when 87.5\% of the total rows are skipped at 2200 fps (\figref{akra_patns} (a)), and are essentially invisible when up to 95.7\% of the total rows are skipped at 5988 fps (\figref{akra_patns} (e)). When the kernel-based inpainting methods are used, the major Kikuchi bands readily emerge (\figref{akra_patns} (b) to (d)). This is evidenced by the similarity of both the pattern and the corresponding fast Fourier transform (FFT) between the 2200 fps data and the full-frame data acquired at 281 fps in \figref{akra_patns} (i).  

With increasing spacing between the sampling rows, larger areas need to be in-painted using the signals from a relatively small fraction of detecting pixels. Correspondingly, the box averaging and linear interpolation inpainting methods become less effective in reconstructing the diffraction patterns. Taking \figref{akra_patns} (f) as an example, the box average method assigns values to boxes with a size of 48$\times$48 (2304 pixels) based only on the signals of 96 detecting pixels inside the box, which accounts for 4.1\% of the boxed area. As a result of the large box size, the image appears to contain large square pixels in \figref{akra_patns} (f), not so dissimilar to a binned conventional EBSD detector. The Telea algorithm gives smoother gradients of inpainting, yet the result (\figref{akra_patns} (h)) also contains regions with similar intensities surrounding the active pixels. Both methods do not visually reconstruct the diffraction pattern when 95.7\% of the pixel rows are inactive (yielding a frame rate of 5988 fps) due to the small number of active pixel rows relative to the vast skipped regions. However, it is important to note that the lack of patterns visible by eye does not imply that diffraction information is not encoded in the images. Despite the incredibly sparse sampling of the detector, much of the diffraction information is still preserved in the images, as will be shown in Section \ref{sec:Indexing}.

Among the three inpainting methods tested, the linear interpolation method produces short range vertical streaks (\figref{akra_patns} (c) and (g)), which are visible as a diffuse horizontal band in the FFTs. There does not appear to be a strong difference between the patterns in-painted using the box average method and Telea algorithm \cite{telea2004}. However, the Telea method requires roughly 250$\times$ as much computational time. inpainting using either box averaging or linear interpolation in Python 3 takes about 18 ms on the 2200 fps data using a high performance computing node with 20 Xeon cores, whereas the Telea algorithm implementation in the Python OpenCV package \cite{opencv} takes roughly 4.5 s. The orders of magnitude time difference makes it impractical to use the Telea algorithm to process the tens of thousands to millions of patterns present in typical EBSD scans (40,000 patterns in our small 200$\times$200 $\mu$m scan at 1 $\mu$m step size).

        \subsubsection{Indexing results} \label{sec:Indexing}

\begin{figure*}[htb!]
\centering
\includegraphics[width=0.9\textwidth]{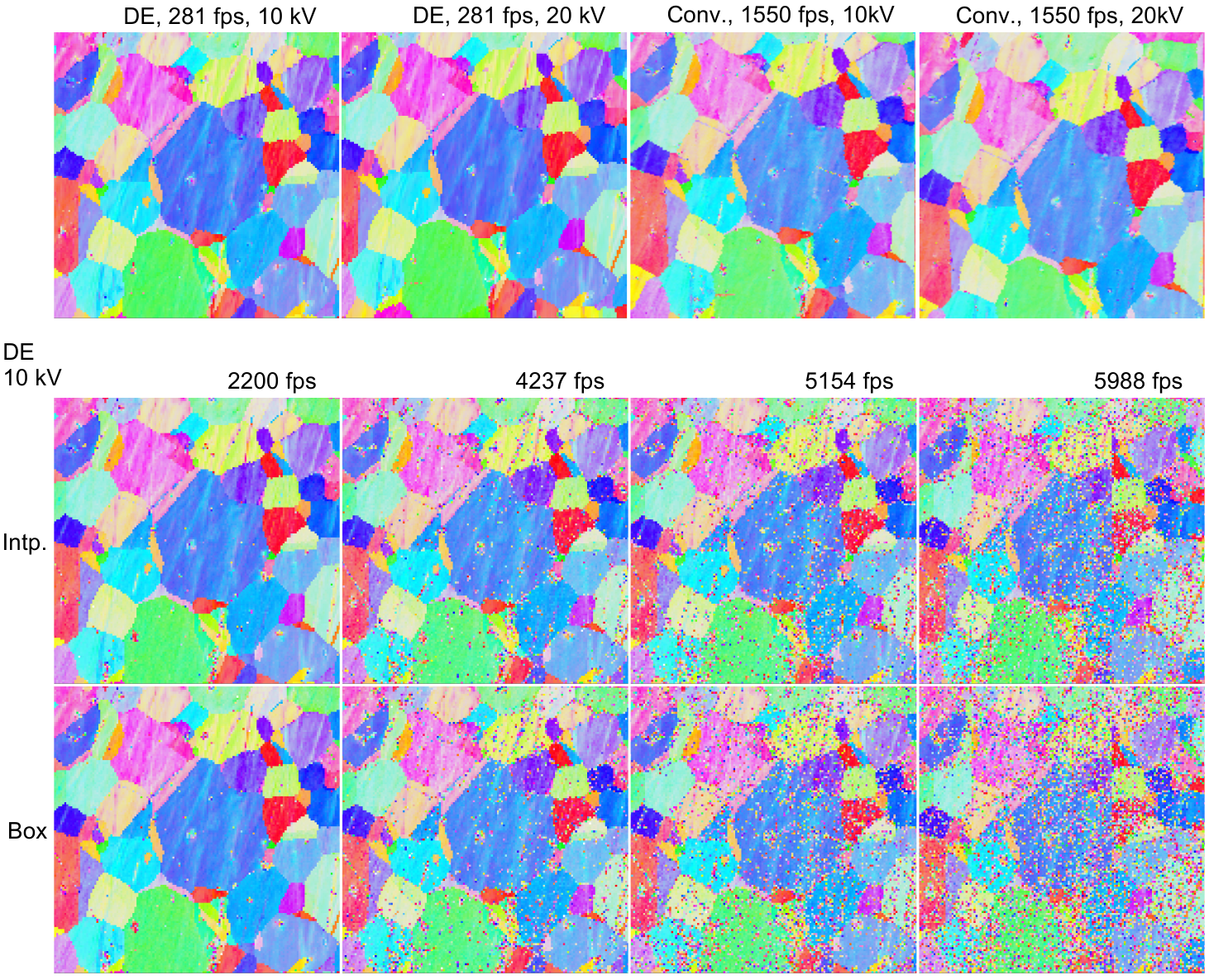}
\caption{Indexed orientation maps of the data acquired using a conventional EBSD detector and the DE-SEMCam, respectively, at 10 keV and 20 keV accelerating voltages and a 32 nA beam current. The scan speed is shown as frames per second (fps). For the fast-scan data on the DE-SEMCam at 2200, 4237, 5154 and 5988 fps, the patterns are processed by inpainting using either linear interpolation (labelled "Intp.") or box average (labelled "Box").}
\label{ipfs}
\end{figure*}

The ultimate measure of the fast-scan EBSD and the post-collection inpainting is the accuracy of indexing. We apply spherical indexing (EMSphInx) to: (1) the data from the DE-SEMCam operating at full-frame at 281 fps, (2) the data from a conventional detector operating at full-frame and increased speed by binning, and (3) the fast-scan data from the DE-SEMCam operating in AKRA mode at four enhanced frame rates (2200, 4237, 5154 and 5988 fps). For consistency, all of the scans were performed at both 10 keV and 20 keV accelerating voltage with a beam current of 32 nA. For the fast-scan data from the DE-SEMCam, the results from two inpainting methods, linear interpolation and box average, are analyzed and compared.

Representative indexing results are presented in \figref{ipfs} in the form of inverse pole figure (IPF) orientation maps. They are arranged in order of increasing pattern collection rate, which was achieved by increasing either the binning size on the conventional EBSD detector or the AKRA sampling step on the DE-SEMCam. The full-frame DE-SEMCam data set at 281 fps represents the best pattern quality, and the 16$\times$16 binned conventional data set at 1550 fps represents the highest speed on the conventional detector. For both conditions, the 10 keV and 20 keV IPF maps are presented in \figref{ipfs}. Qualitative inspection of the IPF maps of the full-frame DE-SEMCam data sets (281 fps) show that the 10 keV and 20 keV patterns are well-indexed. With increasing frame rate, the IPF maps show more misindexed points, arising from the more sparsely sampled diffraction patterns in the AKRA images (\figref{akra_patns}). Also, inpainting using the interpolation method results in fewer misindexed points than those using the box average. 

\begin{figure*}[htb]
\centering
\includegraphics[width=1\textwidth]{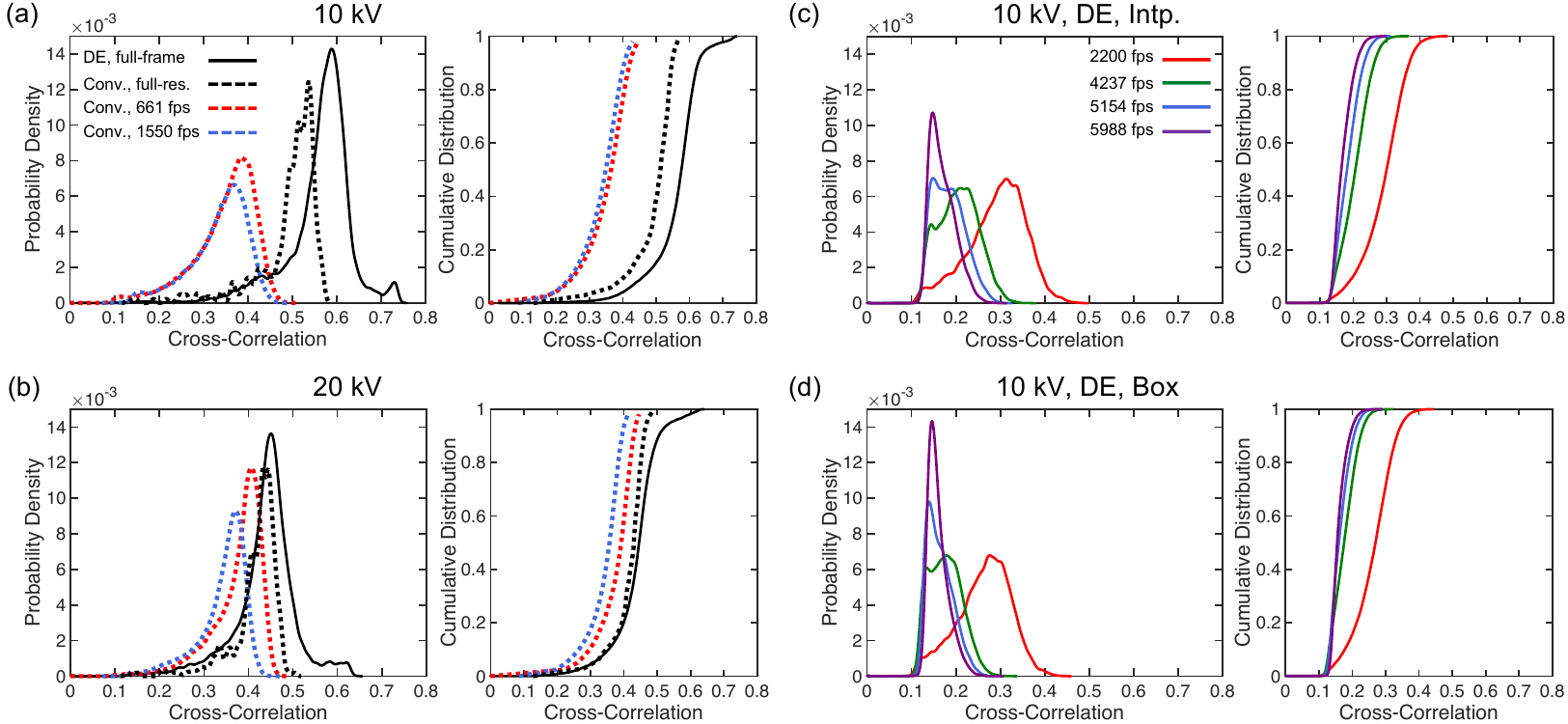}
\caption{The distribution of the XC values for the data acquired using the DE-SEMCam and conventional EBSD detector. Full-frame and full-resolution scans collected with each detector and two higher speed scans collected with the conventional detector using binning modes at 10 keV and 20 keV. For the fast-scan data on the DE-SEMCam at 10 keV, the results from the two inpainting methods, interpolation (Intp.) and box average (Box), are presented in (c) and (d).}
\label{xc}
\end{figure*}

The cross-correlation (XC) value in EMSphInx is used to quantitatively evaluate a confidence metric in indexing. As introduced in section \sectionref{sec:Methods:XC}, a large XC value indicates that the determined orientation has a higher similarity between the experimental pattern and the spherical master pattern, which then implies higher certainty of indexing. The distribution of the XC values for all the points in the scan area is presented in the form of a probability density estimate \cite{matlab_PDF} and cumulative distribution in \figref{xc}. The full-frame data from the DE-SEMCam has the highest XC values among all the scans, most likely due to the high pattern quality that manifests as accurate band position and high image contrast (\figref{highres_pat}). The data acquired at 10 keV have higher XC values than that at 20 keV, which may be linked to the decreased detected band contrast on the DE-SEMCam at 20 keV as shown in \figref{highres_pat} (h). The performance of the conventional EBSD detector is relatively constant at 10 keV and 20 keV (\figref{xc} (a) and (b)). The binning size, namely 4$\times$4 (661 fps) or 16$\times$16 (1550 fps), appears to have a minor influence on the XC values for this Ni sample at relatively high beam current, which is consistent with the report \cite{hansen2017} that the confidence index used in the EDAX software to describe indexing certainty does not appreciably deteriorate at high levels of binning. The XC distributions shift to lower values and become more tightly distributed with increasing scan speed (\figref{xc} (c) and (d)). The interpolation-based inpainting method results in slightly higher XC values than the box average, in agreement with the qualitative observations of the orientation maps in \figref{ipfs}.   

The XC values of the fast-scan data reduce in comparison to the full-frame data, as expected when considering the impressively small active pixel area fraction and short acquisition times (\tableref{t_akra}). In the case of the 2200 fps data set, 28 rows of pixels are in-painted between the active pixel kernel rows. In the 5988 fps data set, 92 rows are in-painted between active pixel kernel rows. When performing whole pattern cross-correlation during the indexing process, the in-painted pixels would result in lower similarity with the theoretical master pattern than the active pixels. At larger sampling step sizes, there are a higher fraction of in-painted pixels, resulting in the cross-correlation value of the pattern to shift toward lower values. On the other hand, a low XC value does not necessarily mean a lack of accuracy in the indexed orientation. The solution may still be correct, because it has the highest cross-correlation value among the possible solutions. Nevertheless, it is notable that the highest frame rates studied here still present values of XC that lead to indexable patterns, in spite of diffraction patterns that appear entirely as noise (\figref{akra_patns}).  This striking result suggests that sufficient correlated diffraction information is encoded in the high frame rate patterns to enable high speed mapping.

        \subsubsection{Error analysis of indexing}

\begin{figure*}[htb]
\centering
\includegraphics[width=1\textwidth]{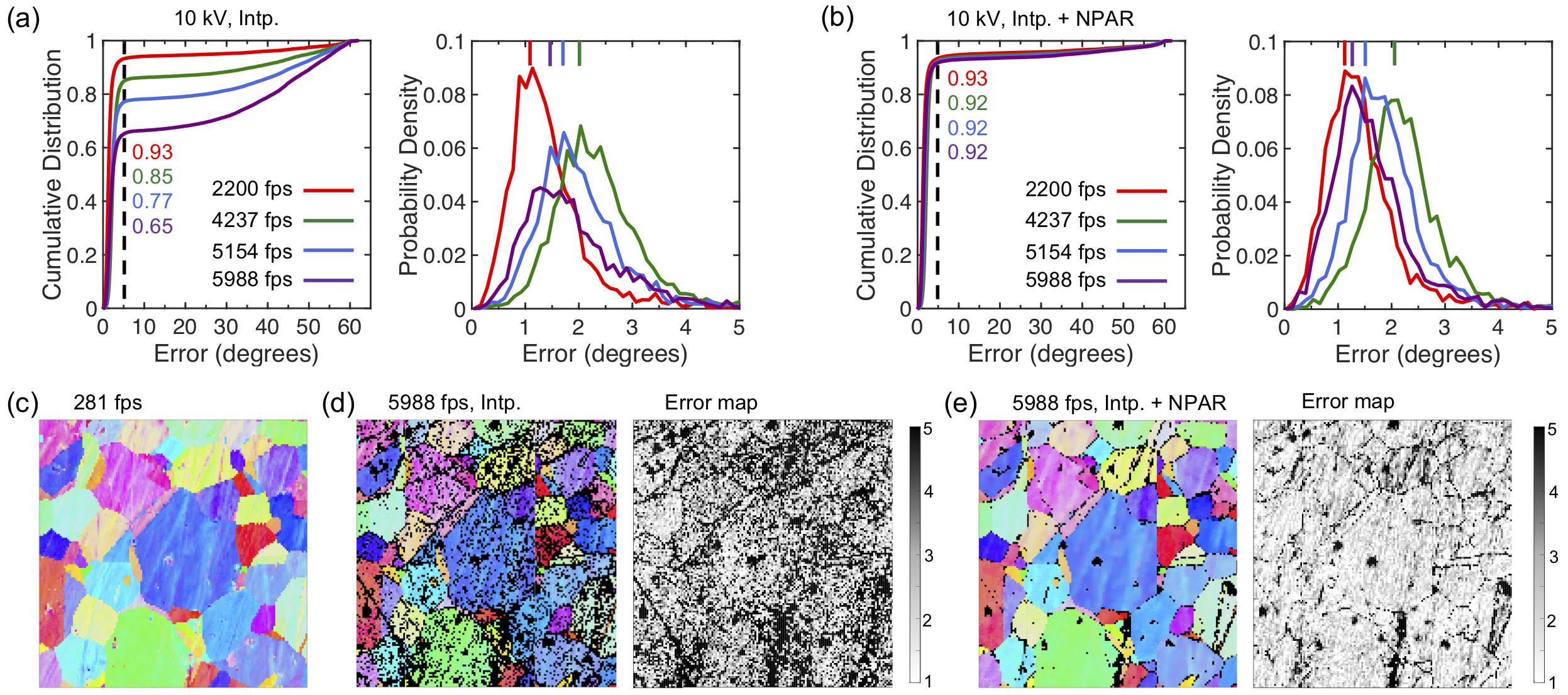}
\caption{Distribution of the orientation error of the fast-scan data at 10 keV with respect to the reference orientations of the 281 fps full-frame data in (c). (a) The fast-scan patterns are processed by inpainting using linear interpolation. (b) The patterns are additionally enhanced by the NPAR \cite{wright2015}. The dashed lines in the cumulative distribution plots are at 5$^\circ$, and the corresponding percentages for each curve are annotated. The vertical bars in the kernel density estimate plots indicate the position of the peak probability. (d) and (e) show the orientation map and error map of the 5988 fps data processed only by inpainting and by a combination of inpainting and NPAR. In the orientation maps, points that have an error larger than $5^\circ$ are colored by black. The error maps are color-coded from 1$^\circ$ to 5$^\circ$.}
\label{misorientation}
\end{figure*}

In the following analysis, the full-frame data (281 fps) from the DE-SEMCam is assumed as the ground truth. The indexing error is calculated using the difference in angles  between the indexed orientations of the AKRA fast-scan data and the ground truth orientations. These angular differences are calculated on a point-by-point basis over the same scan region of the fast-scan data and the full-frame data, so that a distribution of angular difference is obtained for every data set. Since the DE-SEMCam is optimized at approximately 10 keV (i.e. more satisfactory indexing is achieved for the 10 keV DE-SEMCam data using the interpolation inpainting method), the error analysis was performed on the four AKRA data sets in this group. The distribution of the angular error is shown in \figref{misorientation} (a). The vast majority of the indexed orientations are within 5$^\circ$ angular error with respect to the reference and the error peaks around 1$^\circ$ or 2$^\circ$. The rest of the points have an error larger than $30^\circ$, i.e. distinctly different from the reference. These can be ascribed as incorrectly indexed points. As shown in \figref{misorientation} (d), most of these incorrectly indexed points are at grain boundaries (GB) or regions of poor sample preparation quality, e.g. surface contamination or mechanical polishing induced scratches which are also present in the reference map in \figref{misorientation} (c). The grain interiors also periodically contain isolated points of incorrect orientations, likely due to mechanical preparation. 

Whereas the low XC values of the fast-scan data are due to a small number of active detecting pixels, the high sensitivity of the sensing elements guarantees that orientations are reliably determined. The range of the error up to 5$^\circ$ originates from the large spans of inactive pixel rows between the active pixel rows, and therefore the lack of spatial sensitivity for determining the exact Kikuchi band locations. On the other hand, the correctly indexed points (those around 1$^\circ$ or 2$^\circ$ error) provide a basis to apply additional post-collection processing techniques to enhance the indexing, including using neighbor pattern averaging and re-indexing (NPAR) \cite{wright2015}. When extending to the 2nd nearest neighbors (8 neighbors around 1 point in a square scan grid), NPAR yields a high percentage of points ($\sim$93\%) whose re-indexed orientations are within 5$^\circ$ orientation difference from the reference for all the fast-scan data (\figref{misorientation} (b)). Most of the grain interior are correctly indexed (\figref{misorientation} (e)), leaving the incorrectly indexed points almost only at grain boundaries. A recent method using non-local averaging of patterns (NLPAR) \cite{brewick2019} reports to further improve the indexing results, with less or no influence on the spatial resolution of the scan.

\section{Discussion}

The excellent EBSD pattern quality and fast frame rate performance of the DE-SEMCam used here are largely due to the following factors: high electron detection efficiency at the low operating voltages in SEMs, small pixel size/high pixel count of the monolithic active pixel sensor, and the hardware interface that allows sparse sampling of the detector pixel rows. Compared to other MAPS sensor designs, the custom DDD\textsuperscript{\textregistered} sensor in the DE-SEMCam uniquely includes on-chip correlated double sampling (CDS), which is designed to reduce noise in every frame read out by the detector. Examination of frames from the camera in the absence of an electron beam showed that the root-mean-square (RMS) noise was $\sim$0.92 counts. In comparison, we observed that the most-probable intensity on the detector for a 10 keV electron was $\sim$86 counts, corresponding to a very high single-electron signal-to-noise ratio (SNR) of $\sim$93 for operation at 10 kV. This high detection sensitive is critical for operation with low-dose and/or high speed.

The small pixel size gives unprecedented angular resolution over the detector area, namely 0.58 mrad at full resolution. These high angular resolutions are attainable with relatively long acquisition times, e.g. 1 s as in \figref{highres_pat}, which can be employed for structure refinement with the electron beam dwelling in a spot mode or by scanning over a small number of points on the specimen. On the other hand, at the highest full-frame speed of 281 fps, which is suitable for large area EBSD mapping, the patterns are exceptional for indexing compared to full-frame patterns collected with an indirect phosphor-based conventional EBSD detector. The high XC values of the full-frame data acquired on the DE-SEMCam shown in \figref{xc} (a) and (b) corroborates the close similarity of the experimentally collected EBSD patterns to the dynamically calculated master patterns. Depending on the specific material and phenomena of interest, a frame rate up to 281 fps could be adopted in practical applications to provide both a highly desirable pattern resolution, quality, and acceptable scanning speed. This is likely of interest for applications such as HR-EBSD where the EBSD detector resolution is of critical importance. 

As an additional benefit, the EBSD scan speed is greatly boosted when the detector operates in the sparse-sampling AKRA mode to read and save data from selected pixels rows on the detector. The vertically-compressed AKRA images are then post-processed using standard inpainting methods. The primary consequence of the decreased fraction of active pixels is a decrease in the XC value from spherical cross-correlation (e.g. in \figref{xc} (b)). The effect of the short detector exposure times present in the fast AKRA modes is a reduced electron dose on the active pixels. Even with the high detection efficiency of the DE-SEMCam, the low electron sampling statistics over the short exposure time makes inpainting of the vast inactive pixel regions challenging. As shown in \figref{akra_patns} (f) to (h), when as many as 92 rows of inactive pixels exist between 4 active pixel rows, standard inpainting methods become less effective at interpolating the diffraction information between the active rows. The qualitative absence of Kikuchi bands by the human eye does not imply that real diffraction information in the form of correlated noise is absent from the patterns; on the contrary, patterns remain indexable, albeit with a decreasing number of patterns that can be accurately indexed with increased sparsity of active pixel rows, as shown in \figref{misorientation} (a). The small error in the indexed orientations of the majority of the patterns suggests that inpainting introduces only a minor modification to the orientation information carried in the data. Furthermore, a potentially fruitful path forward would be to modify the indexing algorithms to take into account the anisotropic pixel sampling densities, removing the inpainting requirement altogether. For example, Hough-transform based indexing could be modified to account for the non-isotropic aspect ratio of AKRA data. As another example, dictionary indexing could either be modified to create dictionaries with sampling that matches the AKRA readout or to only use AKRA readout rows when performing cross-correlations.

While the indexing of the fast-scan data acquired in AKRA mode can be further improved by applying NPAR, there appears to be a limit for improvement in this vein. The error distributions of all the fast-scan data become virtually the same after NPAR (\figref{misorientation} (b)), with almost no change for the 2200 fps data. The limit of 93\% indexed points is set by the fraction of the points at grain boundaries and regions with poor surface quality that inherently exist in the sample, as shown in \figref{misorientation} (e). In other words, these effects depend primarily on the sample and not on the performance of the detector. The persistent error around 1$^\circ$ and 2$^\circ$ that cannot be further reduced by NPAR is likely linked to the size of the skipped regions. When applying NPAR on a data set, a pattern at each scan point is re-calculated as the pattern averaged with its nearest neighbors. This can be considered as averaging multiple spatially co-located diffraction patterns, and hence is analogous to increasing the amount of signal (via exposure time or increased beam current) gathered to form a pattern. However, the fundamental limit is that large inactive pixel regions exist between the active regions when using the AKRA modes, producing angular uncertainty in the Kikuchi band location due to interpolation. The patterns acquired at high frame rates benefit the most from NPAR processing, because Kikuchi bands that were not restored by inpainting in a single image are now more effectively estimated due to a strengthened active pixel signal. For the patterns that already contain sufficient signal to restore the Kikuchi bands, e.g. 2200 fps data in \figref{akra_patns} (b), pattern averaging only accentuates the Kikuchi bands but does not affect the accuracy of the band position. To achieve 2200 fps, 28 pixel rows are inactive after every 4 actively detecting rows (\tableref{t_akra}). These 28 rows create a region of uncertainty that correspond to 0.93$^\circ$ in the vertical direction of the detector (using the sample-to-detector distance of 22.402 mm), which is expected to be responsible for the 1$^\circ$ error in the indexed orientations. 

The above analyses shed light on future strategies for optimizing data acquisition on the DE-SEMCam for improved indexing fraction and accuracy. First, by making use of the excellent detection efficiency, each active pixel should collect a sufficiently large number of electrons, so that the spatial distribution of the diffracted electrons can be statistically represented among the active pixels.  Subsequently, inpainting can restore the major Kikuchi bands that are required for reliable indexing. The requirement for sufficient detected electrons per pixel can be met by: larger electron beam currents, larger effective pixel sizes of either the actual or binned pixels on the detector, or post-collection pattern averaging. Second, the accuracy of indexing is linked to the size of the skipped region between the detecting pixels, as it defines the angle of uncertainty. The accuracy certainly could be improved by using a smaller sampling step, but at the expense of slower frame rates. In future sensor developments, one option is to distribute the active pixel rows evenly over the detector, rather than using the 4$\times$4 kernel regions. For example, in the 2200 fps data set, presently the active pixels are grouped into 4 consecutive rows before skipping the next 28 rows. If the active pixel rows were distributed evenly over the periodicity of all 32 rows, there would be 7 skipped rows after each 1 active row, corresponding to improved interpolation results. Another option would be to distribute the active rows in an adaptive way. For instance, the active rows can be distributed densely around regions where the electron flux on detector is reduced, and more sparsely where the electron flux is higher. Alternatively, active rows could be clustered around the pattern center where there are more electrons and more sparsely toward the edges of the detector. Obviously, some of these developments require new hardware designs and should be guided by forward modeling EBSD simulations \cite{EMsoft5}. 




\section{Summary}

A direct detector using a monolithic active pixel sensor that is optimized for the accelerating voltages used in scanning electron microscopes is implemented for EBSD applications. High resolution EBSD patterns or high speed pattern acquisition can be achieved, on the same camera, when the detector operates in different modes, allowing a multitude of advanced EBSD applications. At full-frame mode, accurate and sensitive detection of Kikuchi bands is observed from 4 keV to 28 keV accelerating voltages of the primary electron beam, with the optimal contrast occuring between 8-16 keV. The high resolution EBSD patterns are obtainable at low voltages, providing opportunities for imaging of defect structures. With the arbitrary kernel row addressing (AKRA) mode, the pattern acquisition speed leading to indexable patterns is as high as 5988 fps. The high scan speed can greatly facilitate 3D EBSD experiments where reducing mapping time is of upmost importance, experiments where samples are sensitive to electron beam dose, and dynamic in situ mechanical loading experiments where fast sample measurements are critical due to the timescale of sample relaxations. This work advances the use of direct detectors in SEM-based applications such as EBSD, and highlights the role of MAPS direct detectors optimized for EBSD in circumventing the tradeoff between pattern resolution and mapping speed that often faces end users. Exploring the potential benefits of direct detection for other SEM-based diffraction modalities (e.g. 4D-STEM approaches) would be a very interesting future avenue to pursue.


\section*{Acknowledgements}
The research reported here was supported by the Materials Research Science and Engineering Center (MRSEC) at UCSB (MRSEC NSF DMR 1720256) through IRG-1. The research made use of shared facilities of the National Science Foundation (NSF) MRSEC at UC Santa Barbara, DMR-1720256. We also acknowledge support from the Major Research Instrumentation Award NSF DMR-1828628. Use was made of the computational facilities administered by the Center for Scientific Computing at the CNSI and MRL (an NSF MRSEC; DMR-1720256) and purchased through NSF CNS-1725797. MPE and TMP acknowledge a Department of Defense Vannevar Bush Fellowship, grant number N00014-18-1-3031. MDG acknowledges a Department of Defense Vannevar Bush Faculty Fellowship, grant number N00014-16-1-2821. Direct Electron LP acknowledges the support from Department of Energy (Office of Science, Grant DE-SC0018493).

\bibliographystyle{elsarticle-num}
\bibliography{citations}

\end{document}